\documentclass[superscriptaddress,twocolumn,secnumarabic,amssymb, nobibnotes, aps, prd]{revtex4-1}

\usepackage{graphicx,array,multirow}
\usepackage{color}
\usepackage{amsmath}
\usepackage{ulem}

\newcommand{\figwid}{0.947}

\newcommand{\fshadron}{\textrm{X}}

\newcommand{\nucleus}{\textrm{A}}
\newcommand{\nucleusM}{M_\nucleus}
\newcommand{\nucleusX}{M_\fshadron}

\newcommand{\proton}{\textrm{p}}

\newcommand{\protonEb}{\langle \epsilon\rangle_\proton}

\newcommand{\neutron}{\textrm{n}}

\newcommand{\neutronEb}{\langle \epsilon\rangle_\neutron}

\newcommand{\pnM}{M_{\proton/\neutron}}
\newcommand{\pnEb}{\langle \epsilon\rangle_{\proton/\neutron}}

\newcommand{\nucleon}{\textrm{N}}

\newcommand{\baryon}{\widetilde{\textrm{N}}}
\newcommand{\lepton}{\ell}

\newcommand{\dat}{\delta\alpha_\textrm{T}}
\newcommand{\pn}{p_\nucleon}

\newcommand{\dptv}{\delta\vec{p}_\textrm{T}}
\newcommand{\dpt}{\delta p_\textrm{T}}

\newcommand{\ptnv}{\vec{p}_\textrm{T}^{\,\baryon}}
\newcommand{\pln}{p_\textrm{L}^{\,\baryon}}

\newcommand{\ptlv}{\vec{p}_\textrm{T}^{\,\lepton}}
\newcommand{\ptl}{p_\textrm{T}^{\,\lepton}}
\newcommand{\pll}{k_\textrm{L}^{\,\lepton}}

\newcommand{\leptonE}{E_\lepton}
\newcommand{\baryonE}{E_{\baryon}}

\newcommand{\nuzeropi}{\nu 0\pi N\proton}
\newcommand{\nuonepi}{\nu 1\pi N\proton}
\newcommand{\nubarzeropi}{\bar{\nu} 0\pi N\proton}
\newcommand{\nubaronepi}{\bar{\nu} 1\pi N\proton}
\newcommand{\genie}{\textsc{GENIE}}
\newcommand{\gevc}{\textrm{GeV}/\textit{c}}
\newcommand{\mevc}{\textrm{MeV}/\textit{c}}
\newcommand{\gev}{\textrm{GeV}}

\begin{document}

\title{Identification of nuclear effects in neutrino and antineutrino interactions on nuclei using generalized final-state correlations }%

\author{Xianguo Lu}
\email{Xianguo.Lu@physics.ox.ac.uk}
\affiliation{Department of Physics, University of Oxford, Oxford, OX1 3PU, United Kingdom}
\author{Jan T. Sobczyk}%
\email{jsobczyk@ift.uni.wroc.pl}
\affiliation{Institute of Theoretical Physics, University of Wroc\l aw, pl. M. Borna 9,
50-204, Wroc\l aw, Poland}
\date{\today}%

\newcommand{\edit}[1]{{#1}}
\newcommand{\editt}[1]{{#1}}

\newcommand{\edittt}[1]{{\color{green} #1}}

\begin{abstract}
In the study of neutrino and antineutrino interactions in the GeV regime, kinematic imbalances of the final-state particles have sensitivities to  different nuclear effects. Previous ideas based on neutrino quasielastic interactions [Phys. Rev. C94, 015503 (2016); Phys. Rev. C95, 065501 (2017)] are now generalized to antineutrino quasielastic interactions, as well as neutrino and antineutrino pion productions. Measurements of these generalized final-state correlations could provide unique and direct constraints on the nuclear  response inherently different for neutrinos and antineutrinos and, therefore, delineate effects that could mimic charge-parity violation in neutrino oscillations.

\end{abstract}

\maketitle

\section{Introduction}
\label{sec:intro}

It has been well established that good understanding of (anti)neutrino-nucleus cross sections in the GeV regime is necessary for constraining  systematic errors in long-baseline oscillation experiments \cite{Alvarez-Ruso:2017oui, Betancourt:2018bpu}. The most challenging part of this effort is to describe nuclear effects. 
In the impulse approximation (IA) picture, which has a solid foundation in the few-GeV neutrino energy region, nuclear effects such as the target nucleon initial state (Fermi
momentum and binding energy) and final-state interactions (FSIs) impact how individual scattering is seen in experimental setups.
In particular, it is very important to understand well any imprint of nuclear effects on neutrino and antineutrino scattering that could be misunderstood and taken in the data analysis as a manifestation of charge-parity (CP) violation. 

In addition to nuclear effects,
the few-GeV energy region is complicated because of the overlap of many reaction channels, physics mechanisms, etc. A significant unknown is the contribution from the two-body current [called here also two-particle-two-hole (2p2h)], a subject of many experimental and theoretical studies \cite{Marteau:1999kt, Martini:2010ex, Nieves:2011pp, Gran:2013kda, Gonzalez-Jimenez:2014eqa, VanCuyck:2016fab, VanCuyck:2017wfn, Benhar:2015ula, Lovato:2017cux, Rocco:2018mwt}.

In the available models, events coming from the two-body current mechanism contribute almost entirely to the charged-current (CC) $0\pi$ category, defined as having one charged lepton and no pion in the final state. They come there together with CC quasielastic (QE) events and also with  pion production (RES) followed by absorption inside the nucleus. 

Measurements of muon momentum in CC$0\pi$ events are very important for experiments like T2K, where most of the information about the  oscillation signal comes from detection of the final-state muons only. However, those measurements are not sufficient to put constraints on the amount of two-body current contributions. It is why there is a growing interest in measurements involving final-state protons. Interpretation of such measurements is challenging as it requires a good control of proton FSIs in Monte Carlo (MC) event generators \cite{PDG-MC}. This will become most exigent in liquid argon (LAr) experiments with a low-momentum proton detection threshold \cite{Acciarri:2014gev}. It is known that modeling the low-momentum nucleon in-medium cross section is most uncertain \cite{Li:1993rwa}. Challenges come together with opportunities and one may hope to learn from proton studies something new about nuclear physics.   

Another challenge arising with the era of proton observables is that of the amount of information or organization of the data.
One option is to measure and discuss multidimensional cross sections for muon and proton momenta \cite{Abe:2018pwo}. Another possibility is to look at certain projections defined in such a way that their interpretation is simpler, pointing  to particular details of physics mechanisms that are involved. 

An intuitive way to look at proton observables is through 
single-transverse kinematic imbalances (single-TKIs) \cite{Lu:2015tcr}. What is analyzed are only transverse projections of muon and proton momentum vectors on the plane perpendicular to the neutrino direction (known with a good precision). If CCQE interaction occurred on a nucleon at rest and if FSI effects were absent,  the sum of the muon and proton momentum projections would vanish. Thus any deviation from zero tells us about nucleon Fermi motion, FSI effects, and also about other interaction mechanisms. 

Recently two measurements of single-TKIs  were performed by T2K \cite{Abe:2018pwo} and  MINERvA \cite{Lu:2018stk} experiments.  Final-state particle correlations are nontrivial in the presence of nuclear effects: the characteristic imprints from   Fermi motion and intranuclear momentum transfer (IMT) (that in our nomenclature includes the impact of FSI and 2p2h dynamics) are readily identified in the measured cross sections.  The new variable, \textit{transverse boosting angle} $\dat$ (for  details, see Ref.~\cite{Lu:2015tcr}), preserves most of the Fermi motion isotropy and  measures the strength of IMT.  In the  region $\dat<90^\circ$,  both T2K and MINERvA measurements are consistent, showing a common Fermi motion baseline; in the region $\dat>90^\circ$  they differ strongly as $\dat$ increases---an intriguing feature pointing to the energy dependence of IMT.  In the MINERvA measurement, $\dat$ also separates model predictions with boosting and dragging effects (``acceleration" vs ``deceleration") of the FSI~\cite{Lu:2015tcr, Lu:2018stk}, and therefore is able to isolate the peculiar elastic component of the \genie~hA FSI model~\cite{Andreopoulos:2015wxa}. 

A refinement of the kinematic imbalance studies was proposed in Ref.~\cite{Furmanski:2016wqo}. The basic observation was that in the IA regime the assumption that the interaction mechanism was CCQE and no FSI occurred allows to resolve the kinematics completely once the final-state muon and proton are measured. The  additional piece of information used here comes from the  longitudinal components of the final-state muon and proton momenta. The new proposed observable is {\it emulated nucleon momentum} $\pn$. Its first measurement was done by the MINERvA Collaboration~\cite{Lu:2018stk}. In the data there are three interesting regions in $\pn$. A pronounced peak at $\pn\sim 200~\mevc$ is, according to MC simulations, dominated overwhelmingly by CCQE events without FSI. Then there is a tail region $\pn\gtrsim 400~\mevc$ and the intermediate region in between with a lot of structures allowing for detailed studies of interaction mechanisms and FSI effects. The CCQE peak in $\pn$ shows the neutron momentum distribution  and it may seem surprising that a neutrino measurement with all its limitations allows for a nice visualization of the basic nucleon feature---that of Fermi motion.

Keeping in mind the usefulness of experimental studies of nuclear target reactions with the single-TKI and emulated nucleon momentum observables, we would like to extend this approach to other experimental situations. Several important reactions are discussed in the same theoretical framework. We argue that when put together they provide a powerful source of information about reaction mechanisms and nuclear effects.

Processes to be discussed are the following (the $N$ below stands for ``at least one"):
\begin{align}
\nuzeropi:~\nu\nucleus&\to\lepton^-\proton\fshadron,\label{eq:chanA}\\
\nubarzeropi:~\bar{\nu}\nucleus&\to\lepton^+\proton\fshadron,\label{eq:chanB}\\
\nuonepi:~\nu\nucleus&\to\lepton^-\proton\pi^+\fshadron,\label{eq:chanC}\\
\nubaronepi:~\bar{\nu}\nucleus&\to\lepton^+\proton\pi^-\fshadron,\label{eq:chanD}
\end{align}
where $\fshadron$ is a final-state hadronic system consisting of the nuclear remnant with possible additional knocked-out nucleons but
without mesons. It is assumed that one charged lepton, at least one proton, and for Eqs.~(\ref{eq:chanC}) and~(\ref{eq:chanD}) additionally one charged pion, are detected. These include  major $\nu$/$\bar{\nu}$ interaction channels at current and future accelerator-based neutrino experiments.  The first investigation of the $1\pi$ channels [Eqs.~(\ref{eq:chanC}) and~(\ref{eq:chanD})] using single- and double-TKI~\cite{Lu:2015hea} was presented in Ref.~\cite{Pickering:2016mlu}.

The goal of this paper is to propose experimental probes for surgical diagnostics in nuclear effects in both neutrino and antineutrino CC interactions. The potential of the new observables is illustrated by performing numerical simulations using GiBUU and NuWro generators  with the MINERvA and T2K beam fluxes. The plan of the paper is as follows. In Secs.~\ref{sec:dynamics}-\ref{sec:mc}, we present the signal definitions, the formulas of the generalized final-state correlations, and the simulation details. In Secs.~\ref{sec:results} and~\ref{sec:discussion}, we discuss the model predictions and the implications. 

\section{Underlying interaction dynamics}\label{sec:dynamics}

The underlying interaction dynamics of Eqs.~(\ref{eq:chanA})--(\ref{eq:chanD}) in  IA and neglecting FSI are summarized as follows (to simplify the discussion, we neglect in this section but not in the numerical computations diffractive and higher resonant pion production):
\begin{align}
\nu\ \neutron&\to\lepton^-\ \proton,\label{eq:chanNA}\\
&\textrm{not applicable,}
\label{eq:chanNB}\\
\nu\ \proton&\to\lepton^-\ \Delta^{++}\to\lepton^-\ \proton\ \pi^+,\label{eq:chanNC}\\
\bar{\nu}\ \proton&\to\lepton^+\  \Delta^0\to\lepton^+\ \proton\  \pi^-.\label{eq:chanND}
\end{align}
 The process in Eq.~(\ref{eq:chanB}) is forbidden in IA without FSI due to charge imbalance. 

In 2p2h dynamics  neglecting FSI, additional reaction channels underlying Eqs.~(\ref{eq:chanA}) and~(\ref{eq:chanB}) are 
\begin{align}
\nu\ \neutron\  \nucleon&\to\lepton^-\ \proton\  \nucleon,\label{eq:chan2A}\\
\bar{\nu}\ \proton\  \proton&\to\lepton^+\ \neutron\  \proton,\label{eq:chan2B}
\end{align}
where the above $\nucleon$ stands for  either a proton or a neutron. One can see that Eq.~(\ref{eq:chanB}) becomes possible as a result of the 2p2h process. We disregard pion production in the 2p2h mechanism, about which very little is known.

When FSI sets in, many new scenarios contributing to reactions in Eqs~(\ref{eq:chanA})--(\ref{eq:chanD}) become possible. Most importantly,  
the  pions resulting from primary interactions can be absorbed, and the   nucleons can knock out other nucleons or even pions seen in the final state. 
Among other channels, Eq.~(\ref{eq:chanB}) is  proposed here for its pure nuclear-effect origin. Its unique feature, as shown in following sections,   is the strongly reduced influence from the Fermi motion.

\section{Generalized final-state correlations}

The nuclear target processes in Eqs.~(\ref{eq:chanA})--(\ref{eq:chanD}) can be summarized as
\begin{align}
\label{eq:joint}
\nu/\bar{\nu}+\nucleus&\to\lepton+\baryon+\fshadron,
\end{align}
where  $\baryon$ is a proton  in Eqs.~(\ref{eq:chanA}) and~(\ref{eq:chanB}) and a  $\proton\pi$ pair  in Eqs.~(\ref{eq:chanC}) and~(\ref{eq:chanD}). 
Similarly, in   IA the reactions  in 
 Eqs.~(\ref{eq:chanNA}), (\ref{eq:chanNC}), and (\ref{eq:chanND}) can be summarized as
\begin{align}
\label{eq:jointIA}
\nu/\bar{\nu}+\nucleon&\to\lepton+\baryon.
\end{align}
Accordingly, the definitions of single-TKI given in Ref.~\cite{Lu:2015tcr} are generalized  so that 
\begin{align}
\dptv &= \ptlv + \ptnv \label{eq:dpt}, 
\end{align}
where $\ptlv$ and $\ptnv$ are the transverse momenta of the lepton and $\baryon$, respectively. The  definition of the transverse boosting angle keeps its original form:
\begin{align}
\dat&\equiv\arccos\frac{-\ptlv\cdot\dptv}{\ptl\dpt}.\label{eq:datdef}
\end{align}

Assuming that the target nucleus was at rest and no other particles were knocked out (i.e., $\fshadron$ is the nuclear remnant of mass $\nucleusX$), one can resolve the kinematics of the process following the steps from Ref. \cite{Furmanski:2016wqo}. The result for the longitudinal component of the target nucleon momentum is
\begin{eqnarray}
\label{eq:solution_p}
p_L= &\frac{1}{2}(\nucleusM + \pll + \pln - \leptonE - \baryonE ) \nonumber \\
     &-\frac{\displaystyle \dpt^2 + {\nucleusX}^2}{\displaystyle 2( \nucleusM + \pll + \pln - \leptonE - \baryonE ) }, 
\end{eqnarray}
where $\nucleusM$ is the target nucleus mass, and $\pll$ ($\pln$) and $\leptonE$ ($\baryonE$) are the longitudinal momentum and energy of  $\ell$ ($\baryon$), respectively.

The emulated nucleon momentum is defined as
\begin{eqnarray}
\label{eq:full_p}
\pn\equiv \sqrt{\dptv^{\,2} + {p_L}^2}.
\end{eqnarray}

The value of $\nucleusX$ can be expressed in terms of the target nucleus mass $\nucleusM$ and the  proton/neutron mean excitation  energies $\pnEb$: 
\begin{equation}
\nucleusX=\nucleusM-\pnM+\pnEb,
\end{equation}
where $\pnM$ is proton or neutron mass. The values used in this paper are $\neutronEb=28.7$~MeV and $\protonEb=26.1$~MeV (see Table 8 of Ref.~\cite{Bodek:2018lmc}).

\section{Simulations}\label{sec:mc}
\label{sec:simulations}

\subsection{NuWro}

NuWro \cite{Golan:2012wx} is a versatile MC neutrino event generator  
developed over last 13 years at Wroc\l aw University. It
provides a complete description of neutrino/antineutrino interactions on arbitrary nucleon and nuclear targets in the energy range from $\sim100$~MeV to
$\sim1$~TeV. The basic interaction  modes  on a free-nucleon target are:
\begin{itemize}
\item  CCQE: see Eq.~(\ref{eq:chanNA}), and its neutral current counterpart,
\item RES: covering a region of invariant
hadronic mass $W\leq 1.6$~GeV; the dominant RES process is $\Delta(1232)$-resonance excitation as in Eqs.~(\ref{eq:chanNC}) and~(\ref{eq:chanND}), \item DIS (jargon in the neutrino MC community   for shallow and deep inelastic scattering \cite{PDG-MC}): all the inelastic processes with $W\geq 1.6$~GeV.
\end{itemize}

In the case of neutrino-nucleus scattering, two new
interaction modes are:
\begin{itemize}
\item COH: coherent pion production,
\item MEC: two-body current processes, called also 2p2h. 
\end{itemize}

Neutrino-nucleus CCQE, RES, DIS, and MEC reactions are
modeled as a two-step process; the primary interaction on one
or two nucleons is followed by FSI.

NuWro FSI effects are described by a custom-made semiclassical
intranuclear cascade (INC) model \cite{Golan:2012wx}. It includes pion absorption and charge-exchange reactions treated according to the model of Oset~\textit{et al.} \cite{Salcedo:1987md, Oset:1987re}.  Values of nucleon-nucleon in-medium
cross section are based on the computations from Ref. \cite{Pandharipande:1992zz}.

In this paper we use NuWro configuration 17.09. CCQE is described with the local Fermi gas (LFG) model, and the standard vector and axial form factors  with the  axial mass value of $1.03$~GeV. RPA effects are added following Ref. \cite{Graczyk:2003ru}. RES is based on
$\nucleon$-$\Delta(1232)$ transition 
axial form factors found in  Ref. \cite{Graczyk:2009qm} by a fit to ANL and BNL pion production data. The nonresonant contribution is added incoherently as explained in Ref.~\cite{Juszczak:2005zs}. The nuclear target pion production cross section is reduced due to in-medium self-energy implemented in the
approximate way using results of Ref.~\cite{Sobczyk:2012zj}. Finite $\Delta(1232)$ life-time effects are also included \cite{Golan:2012wx}, as well as realistic angular distributions of pions resulting from $\Delta(1232)$ 
decays \cite{Sobczyk:2014xza}. DIS is based on inclusive neutrino cross-section computations of Bodek and Yang with hadronization modeled using PYTHIA fragmentation routines. MEC is based on the 
Nieves \textit{et al.}
model~\cite{Nieves:2011pp} with a momentum transfer cut
$q\leq 1.2~\gevc$~\cite{Gran:2013kda}. As for the MEC hadronic part a model from Ref.~\cite{Sobczyk:2012ms} is used. It is assumed that in 85\% of MEC events the interaction occurs on a proton-neutron   pair~\cite{Carlson:2001mp, Subedi:2008zz}.

\subsection{GiBUU}

GiBUU~\cite{Buss:2011mx, Gallmeister:2016dnq} is a theoretical model and also an event generator  describing nuclear interactions with nuclei, including photon-, lepton-, hadron-, and nucleus-nucleus reactions, with a consistent treatment of nuclear effects and a sophisticated  kinetic hadronic transport  framework. In these calculations, both the initial- and final-state hadrons    are embedded in  a coordinate- and momentum-dependent potential. The 2017 version is used in this study. 

The target-nucleon momentum is sampled like in the LFG approach, but due to the nuclear potential the bound nucleon has an effective mass. In this approach, inclusion of RPA correlations are not  needed (see Refs.~\cite{Martini:2016eec, Nieves:2017lij, Sobczyk:2017mts}). The axial mass parameter  in the dipole form factor for the quasielastic scattering is set to 1 GeV.

In  the pion production kinematic region ($W<2~\gev$), the vector couplings and transition form factors  are  determined by the MAID analysis~\cite{Drechsel:2007if}. The  axial part for heavier resonances  is determined by Partially Conserved Axial Current (PCAC) arguments and an assumption of a dipole form factor with the axial mass parameter of 1~GeV,  whereas  for  $\Delta(1232)$ the axial part was  obtained by a fit to bubble-chamber data~\cite{Radecky:1981fn, Lalakulich:2010ss}. 
The nonresonant contributions (together with the interference one) are added in an incoherent way.  
Free spectral functions without in-medium corrections are used for the  $\Delta$-resonance~\cite{Mosel:2017ssx}.  GiBUU does not provide predictions for the coherent pion production. Inelastic processes at  $W$ above 2~$\gev$ are   described as DIS by PYTHIA~\cite{Leitner:2008ue}.  

The 2p2h contribution in GiBUU is fully determined by the structure functions $W_1$ and $W_3$. 
By neglecting the longitudinal part of the response, both structure functions are directly related to the  structure functions  measured in electron-nucleus  scattering~\cite{Bosted:2012qc}.
The relative numbers of initial neutron-proton, neutron-neutron, and proton-proton pairs are determined by  combinatorics arguments. 

After primary interactions, final-state particles are transported on-shell 
in phase-space volumes where quantum statistical effects like Pauli blocking are handled. GiBUU allows for an off-shell transport of hadrons but the results do not change much, and therefore  this option is not used in this study. Pion absorption in FSIs is modeled as two- and three-nucleon processes~\cite{Mosel:2017ssx}; charge-exchange reactions  are described in Ref.~\cite{Buss:2006yk}.

\subsection{MINERvA selection criteria}
\label{sec:minerva_criteria}

Predictions are calculated  with the NuMI low-energy beam flux~\cite{Aliaga:2016oaz} on carbon  targets with the following particle selection: 
\begin{itemize}
\item muon
\begin{itemize}
\item $\theta_\mu<20^\circ$ 
\item $1.5<p_\mu<10~\gevc$ 
\end{itemize}
\item proton
\begin{itemize}
\item $\theta_\proton<70^\circ$ 
\item $0.45 <p_\proton<1.2~\gevc$  
\item at least one proton satisfies the above criteria and the most energetic one is selected in the analysis
\end{itemize}
\item charged pion 
\begin{itemize}
\item $\theta_\pi<70^\circ$ 
\item $75<T_\pi<400$~MeV 
\item exactly one charged pion satisfies the above criteria
\end{itemize}
\item no mesons otherwise, 
\end{itemize}
where $p$, $T$, and $\theta$ are the particle momentum, kinetic energy, and   angle with respect to the neutrino direction. These  selection criteria are derived from MINERvA measurements (for example, Refs.~\cite{Lu:2018stk, Mislivec:2017qfz}). In nonmagnetized detectors like the MINERvA scintillator tracker,  particle momentum is determined by  range. The $T$ upper cuts for protons and pions are to remove  particles undergoing secondary interactions in the detector, to guarantee a precise momentum measurement.

\subsection{T2K selection criteria}
\label{sec:t2k_criteria}

Predictions are calculated  with the T2K beam flux~\cite{Abe:2011ks} on carbon targets with the following particle selection:
\begin{itemize}
\item muon
\begin{itemize}
\item $\theta_\mu<126.87^\circ$  ($\cos\theta_\mu>-0.6$)
\item $p_\mu>0.25~\gevc$
\end{itemize}
\item proton
\begin{itemize}
\item $\theta_\proton<66.42^\circ$  ($\cos\theta_\proton>0.4$)
\item $0.45<p_\proton<1~\gevc$
\item at least one proton satisfies the above criteria and the most energetic one is selected in the analysis
\end{itemize}
\item charged pion 
\begin{itemize}
\item $\theta_\pi<70^\circ$ 
\item $75<T_\pi<400$~MeV 
\item exactly one pion satisfies the above criteria
\end{itemize}
\item no mesons otherwise.
\end{itemize}

These selection criteria are derived from T2K measurements (for example, Refs.~\cite{Abe:2018pwo, Abe:2016aoo}). In T2K, the Time Projection Chamber (TPC) of the near detector can provide precise momentum measurements of pions above $T_\pi=400$~MeV. The upper $T_\pi$ cut here is to have a consistent signal definition as  in MINERvA so that the predictions can be compared within similar phase space. 

\section{Results}\label{sec:results}

\begin{figure}[!ht]
\centering
\includegraphics[width=\figwid\columnwidth]{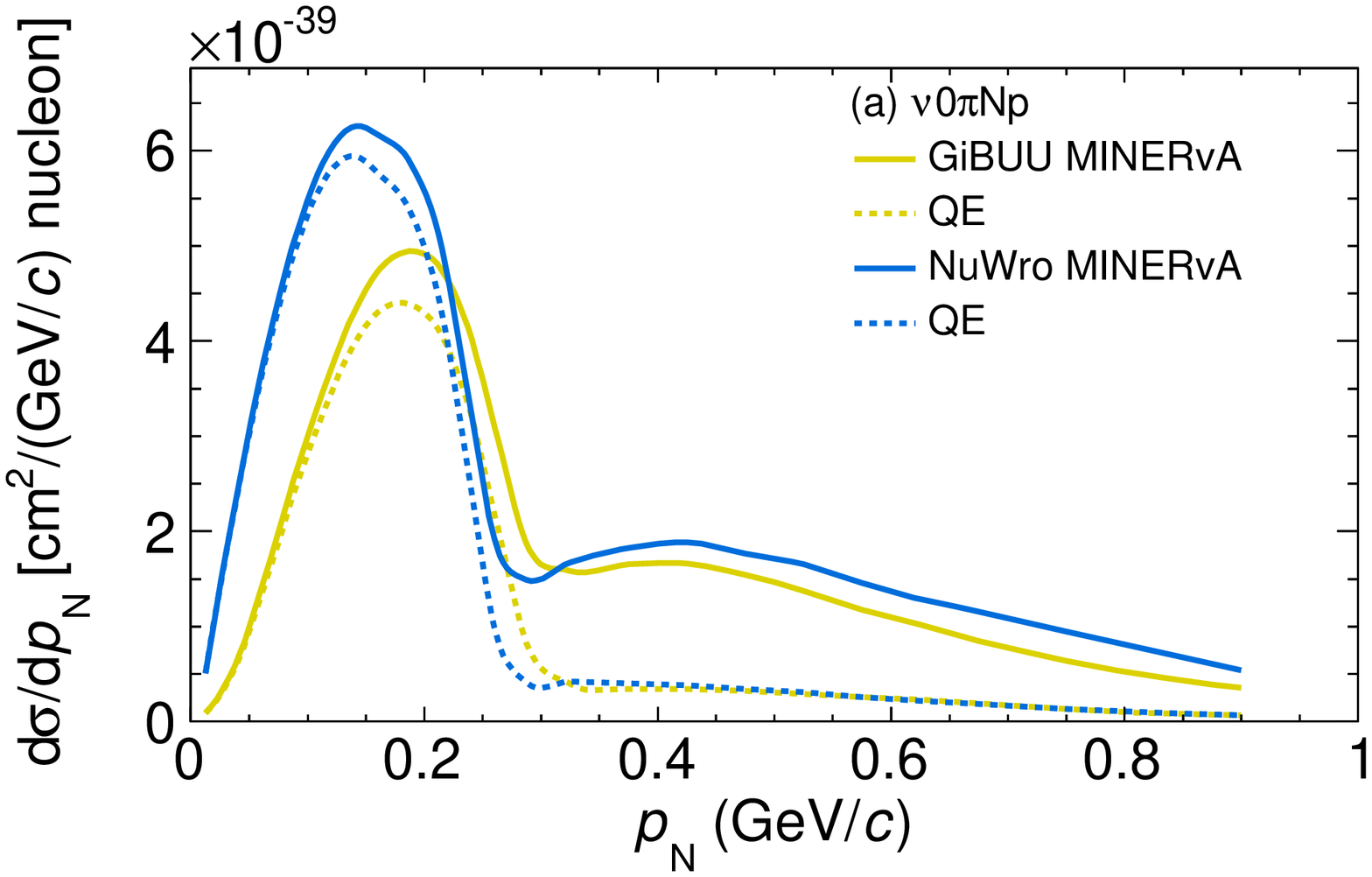}\\
\includegraphics[width=\figwid\columnwidth]{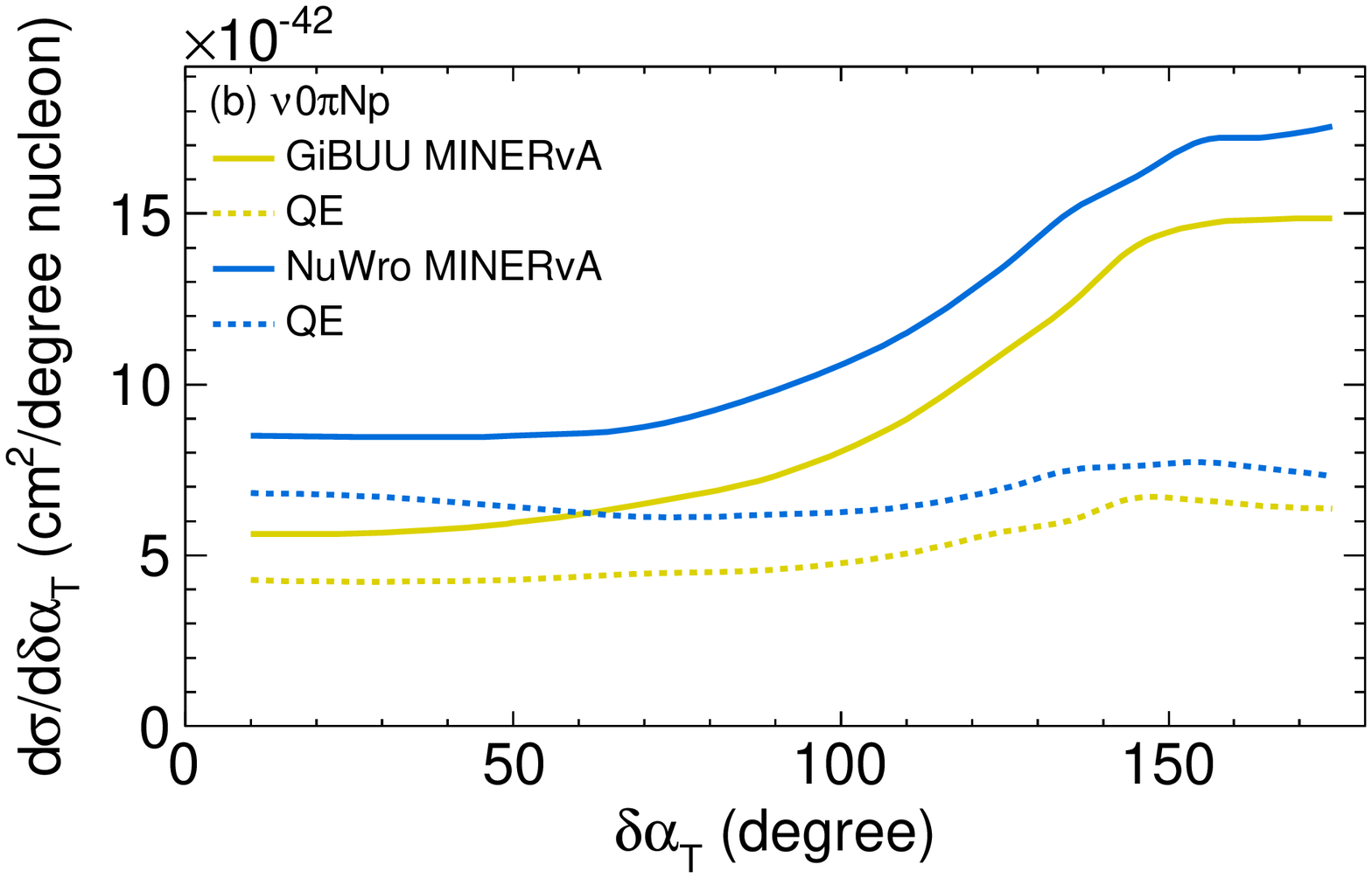}\\
\includegraphics[width=\figwid\columnwidth]{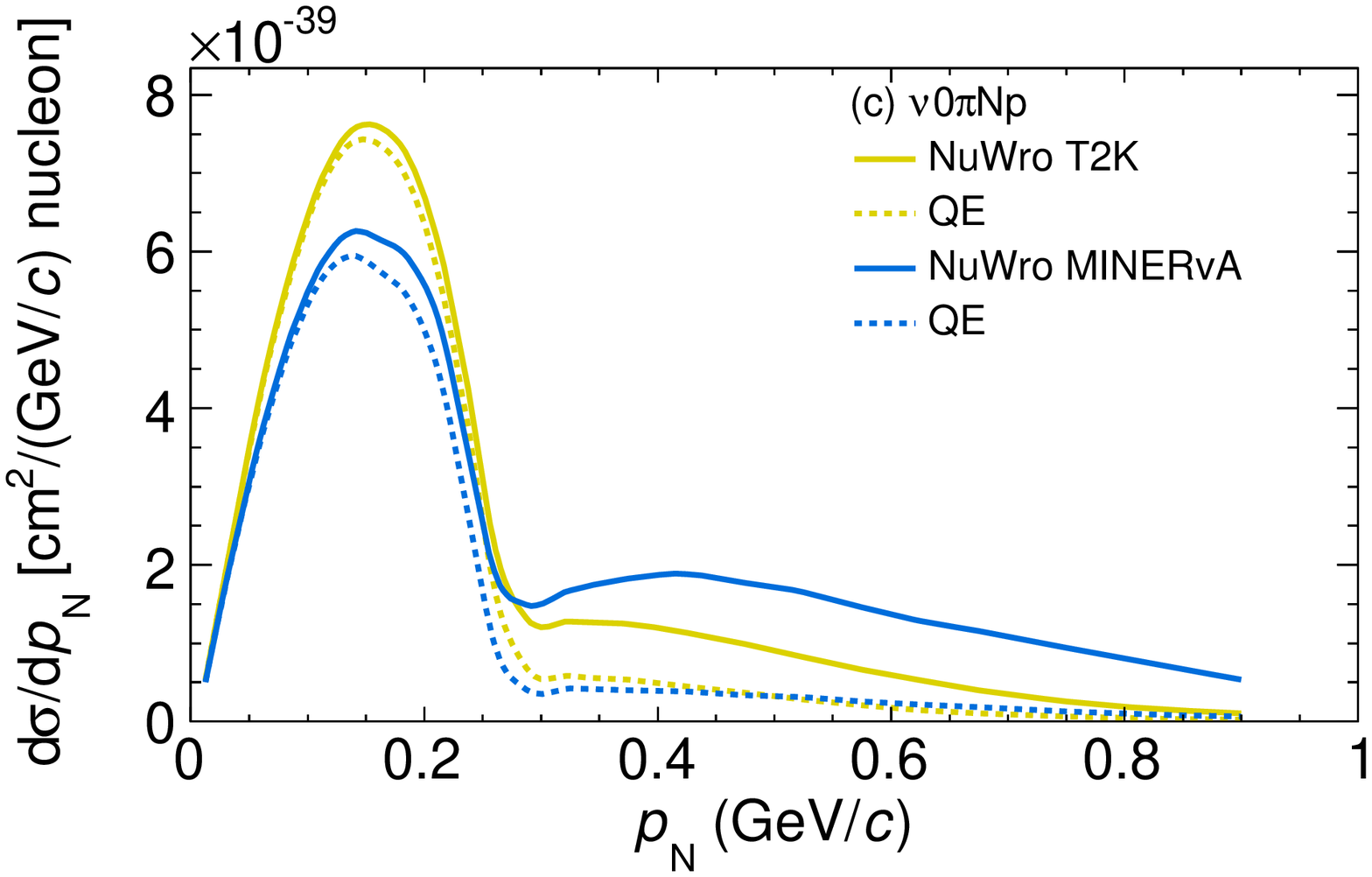}\\
\includegraphics[width=\figwid\columnwidth]{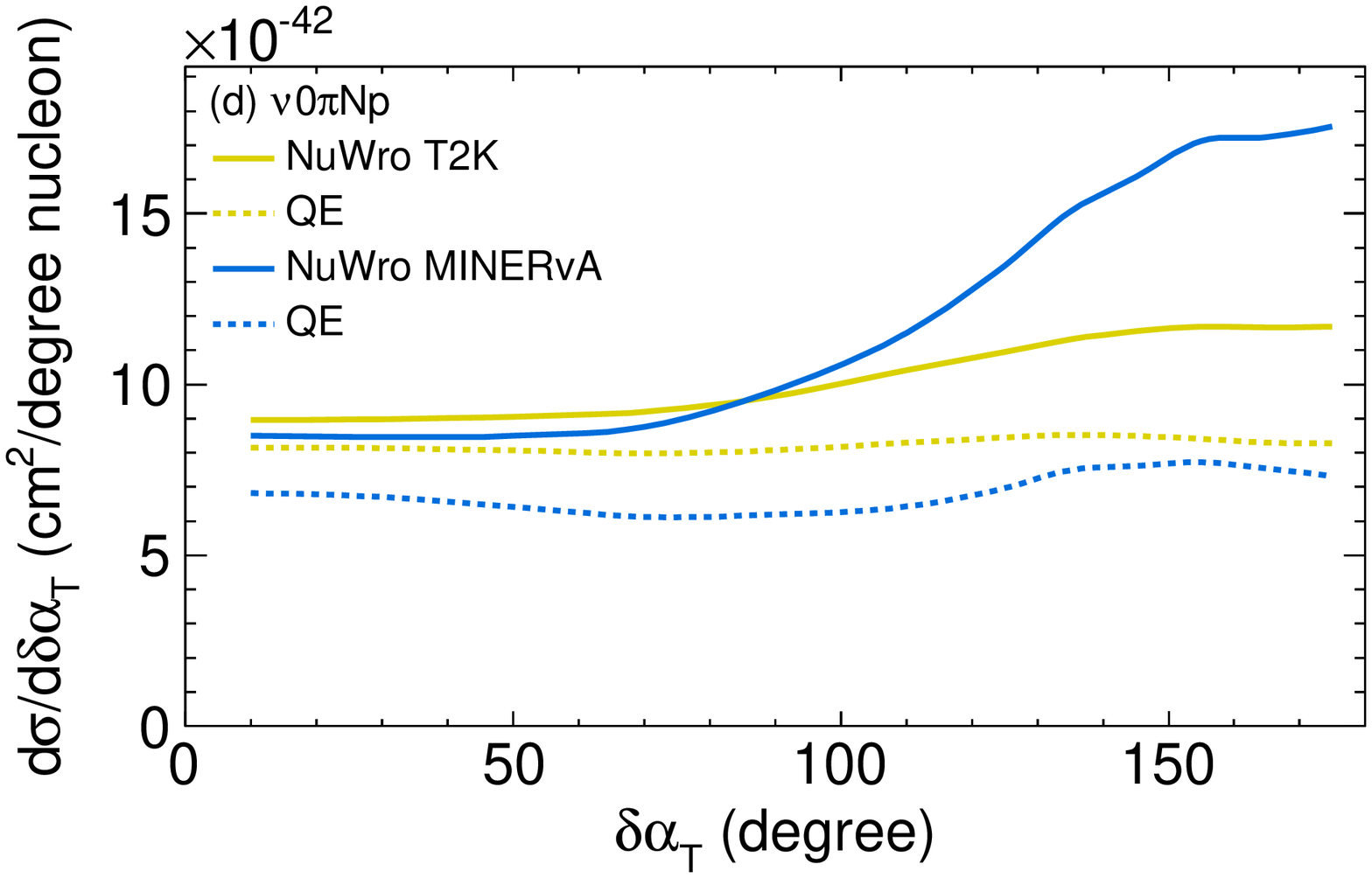}
\caption{GiBUU and NuWro predictions for the $\nuzeropi$ selection. Differential cross sections in $\pn$ and $\dat$, respectively, are compared between the two generators (upper two panels) and between MINERvA and T2K using NuWro (lower two panels). The corresponding QE components are shown.} \label{fig:nuzeropi}
\end{figure}

In Fig.~\ref{fig:nuzeropi}, results for the $\nuzeropi$ selection are shown. The  top two panels present the  results with the MINERvA flux as predicted by GiBUU and NuWro.  The bottom two panels show the NuWro results with both MINERvA and T2K fluxes. The theoretical predictions contain contributions  from several dynamical mechanisms: QE, RES+DIS with pion absorption, and 2p2h. We  show RES+DIS rather than RES and DIS separately because the RES and DIS definitions in NuWro and GiBUU do not match but the sums do. 

As discussed in Ref.~\cite{Furmanski:2016wqo}, $\pn$ is defined in such a way that, for QE events 
where the knocked-out proton does not suffer 
from FSI effects, it is equal to the target-neutron momentum. This explains the peak in the $\pn$ distribution at 150--200~MeV/$c$: it comes from the neutron Fermi motion. 
The  comparison between NuWro and GiBUU  indicates that the initial state  is  modeled   differently.  
It is clear that
with the experimental data it is possible to discriminate between theoretical models (see Refs.~\cite{Lu:2018stk,Dolan:2018zye}). For example, the hole spectral function approach \cite{Benhar:1994hw} as implemented in NuWro provides much better agreement with the data than LFG with RPA corrections (see Ref.~\cite{Lu:2018stk}). For both experiments the shape and position of the peak in the $\pn$ distribution predicted by NuWro are very similar and the difference is mostly in its height (the T2K peak is higher). 
Our understanding is that MINERvA has on average more energetic protons which are removed from the Fermi motion peak by stronger FSI to the right tail.  

For $\dat$, the nonflatness of the distribution of the QE events indicates the strength of the FSI experienced by the knocked-out protons. In the bottom panel we see that  the fraction of non-QE events gradually increases towards the large $\dat$ direction, and at $\dat=180^\circ$ the beam-energy dependence becomes maximal.

\begin{figure}[!ht]
\centering
\includegraphics[width=\figwid\columnwidth]{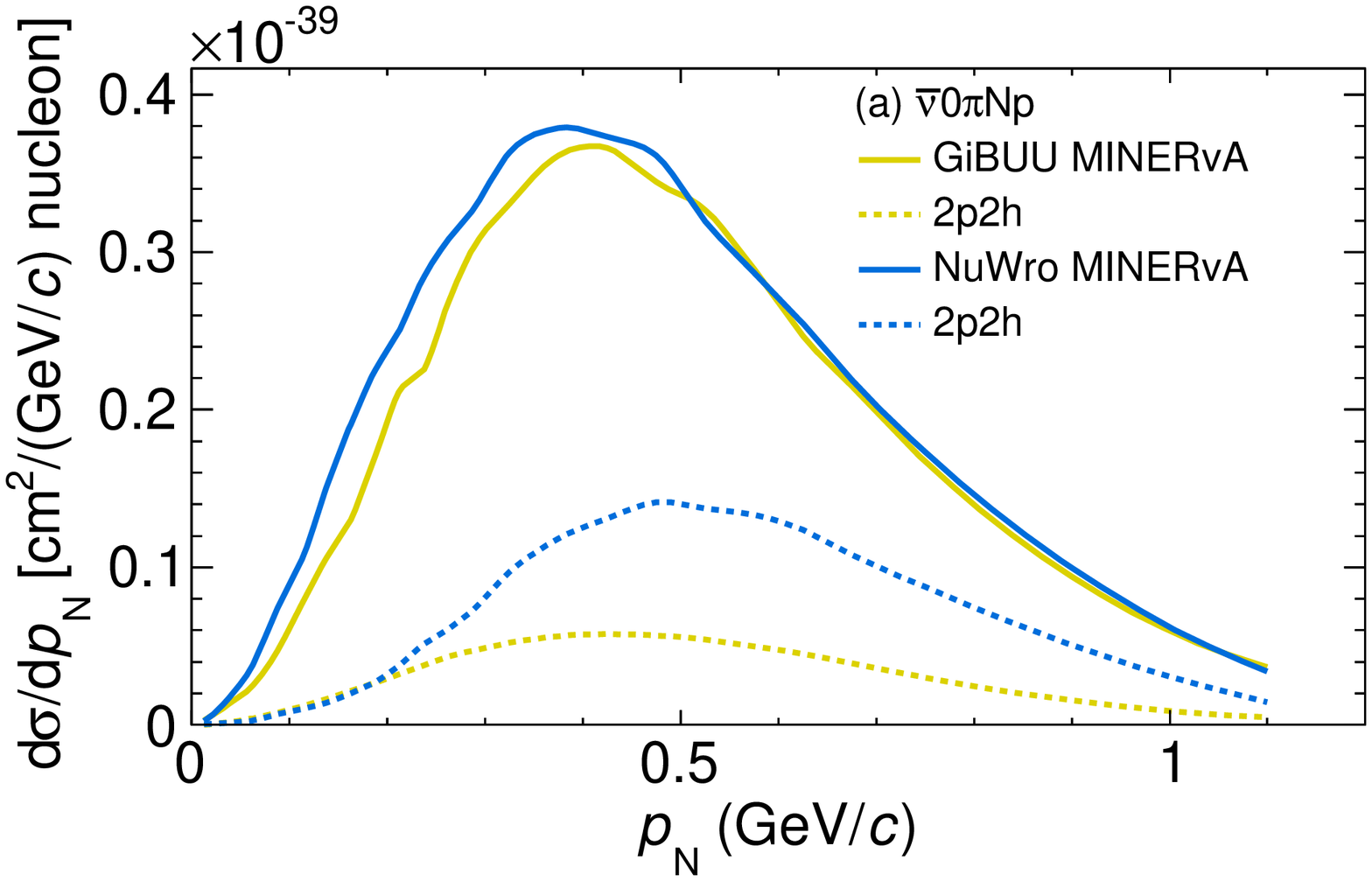}\\
\includegraphics[width=\figwid\columnwidth]{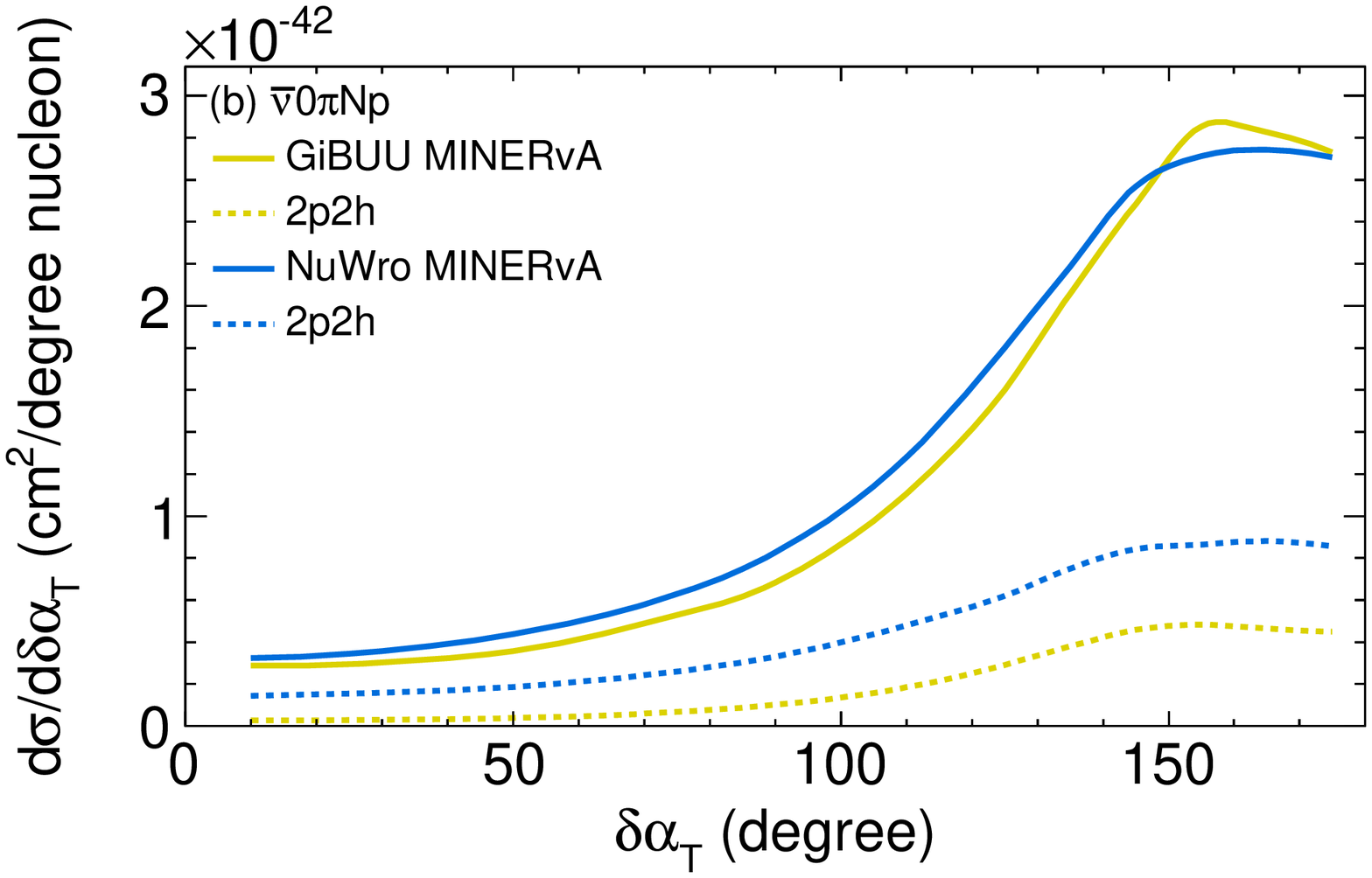}\\
\includegraphics[width=\figwid\columnwidth]{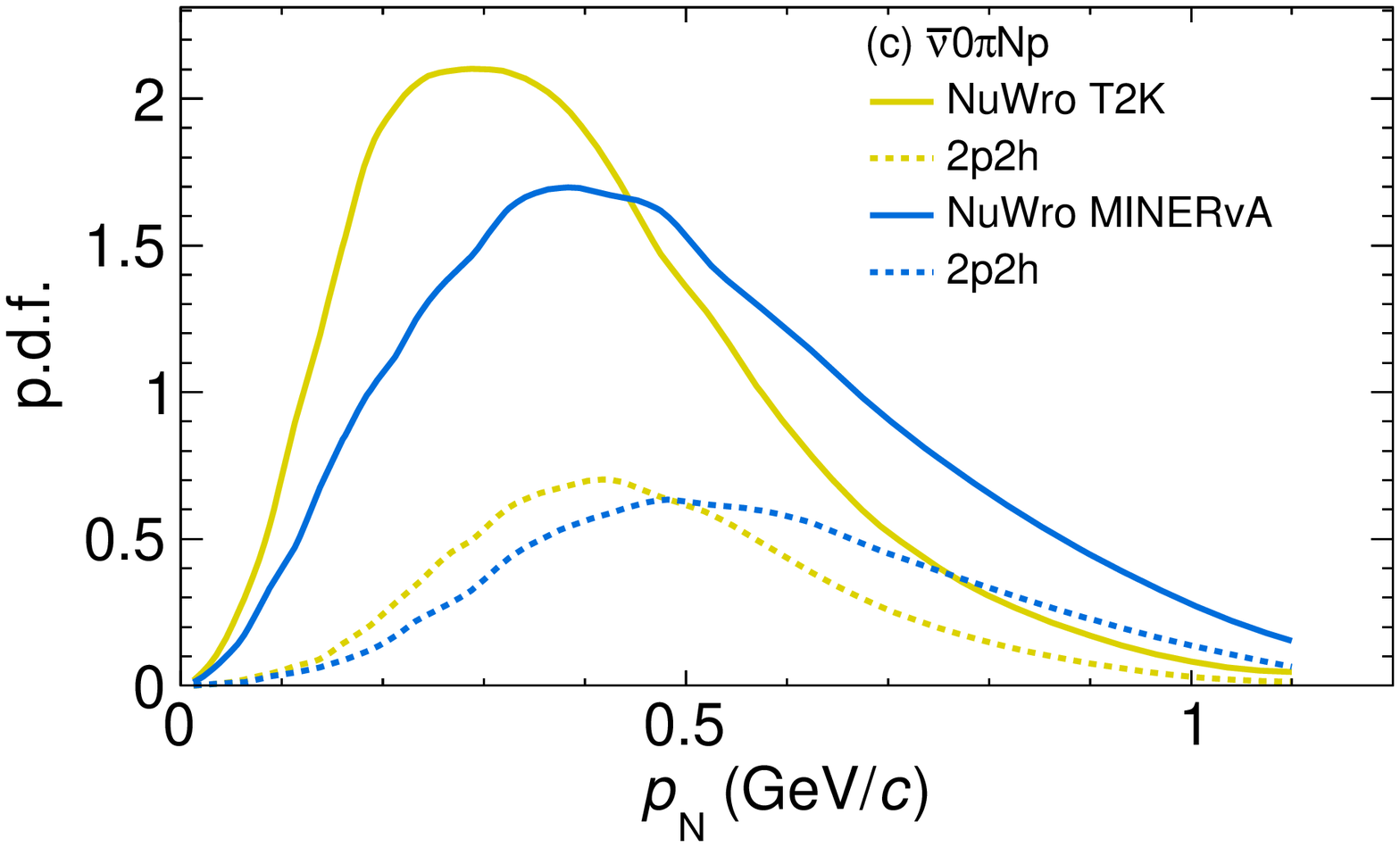}\\
\includegraphics[width=\figwid\columnwidth]{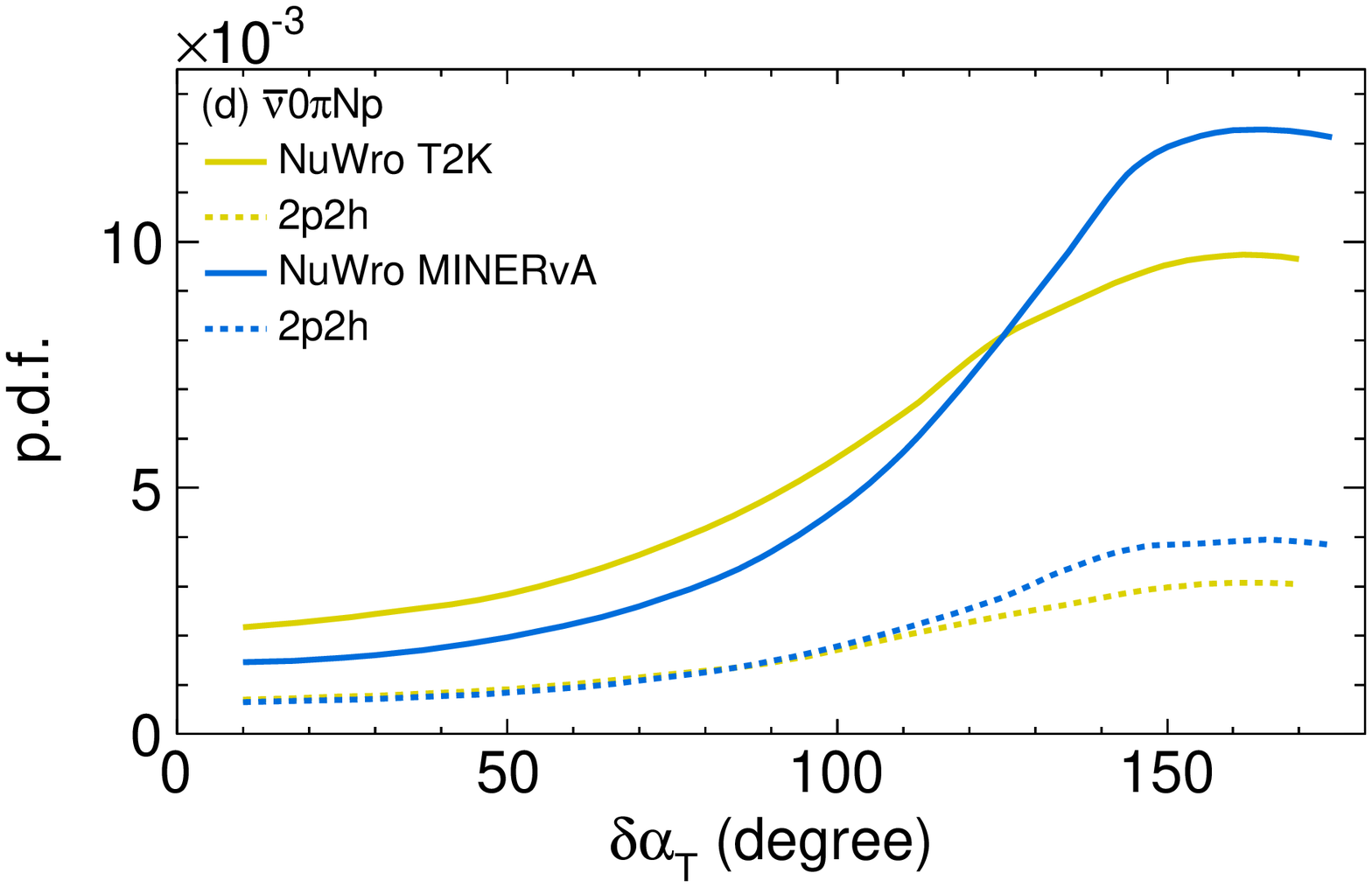}
\caption{Model comparisons for $\nubarzeropi$ in the same layout as in Fig.~\ref{fig:nuzeropi}. The comparisons between MINERvA and T2K in the lower two panels are shown in shape. The 2p2h components are shown.  }\label{fig:nubarzeropi}
\end{figure}

In Fig.~\ref{fig:nubarzeropi} results for the $\nubarzeropi$ selection are shown in the same format as in Fig.~\ref{fig:nuzeropi}.
This channel only includes QE events with charge-exchange nucleon FSI; therefore,  compared to the $\nuzeropi$ channel the dominant Fermi motion peak is absent, and the rise of $\dat$ is much steeper. 
An interesting observation with  this selection is that the  GiBUU and NuWro overall predictions are very similar in shape and normalization,  and yet this agreement turns out to be accidental since individual contributions from interaction modes are quite different. This is illustrated with the 2p2h contributions shown separately. 

$\nuzeropi$ and $\nubarzeropi$  contain complementary information about FSI and 2p2h mechanisms.
For the $\nuzeropi$ selection, QE FSI events are those with quasielastic  proton rescattering. In the $\nubarzeropi$ selection, nucleon charge-exchange FSI is needed for the QE mechanism; the  2p2h contribution comes either from proton-proton initial pairs without FSI or from proton-neutron pairs with charge-exchange FSI. NuWro assumes a much bigger fraction of initial proton-neutron pairs than GiBUU.  We see that the two channels are sensitive to different details of the nucleon FSI and 2p2h mechanisms, and therefore a combined analysis of both channels would help to reveal the full picture of these dynamics.

\begin{figure}[!ht]
\centering
\includegraphics[width=\figwid\columnwidth]{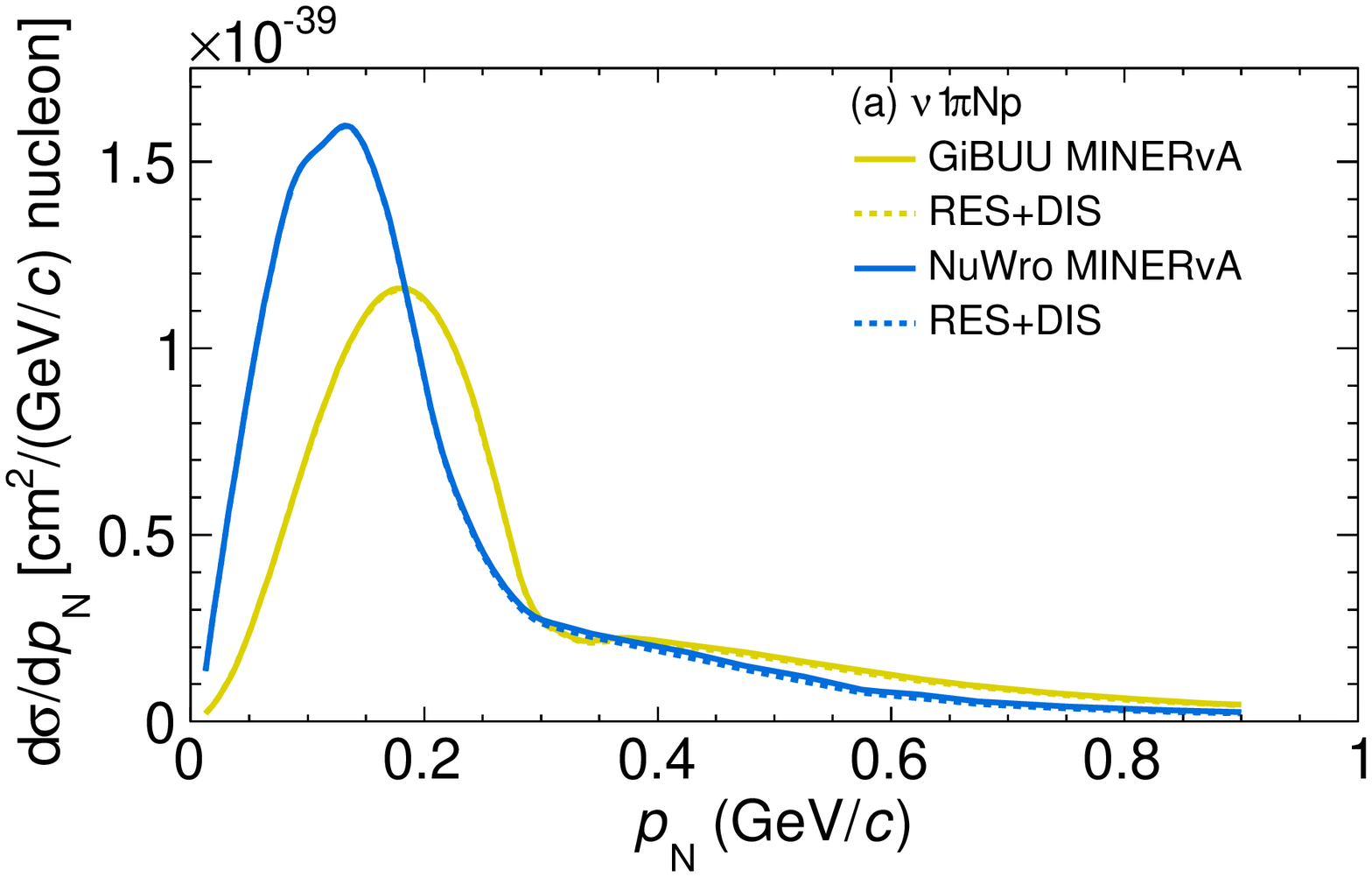}\\
\includegraphics[width=\figwid\columnwidth]{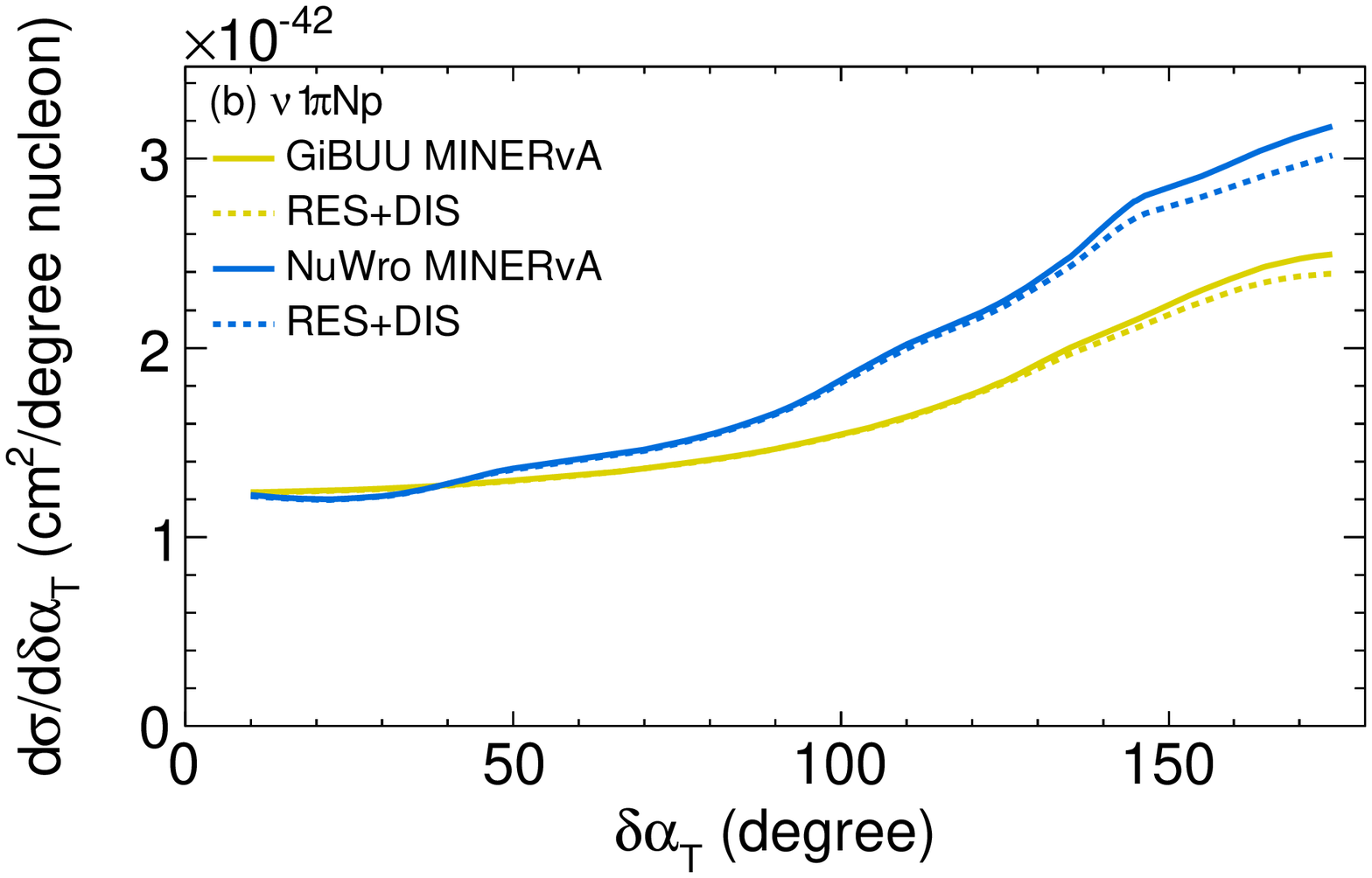}\\
\includegraphics[width=\figwid\columnwidth]{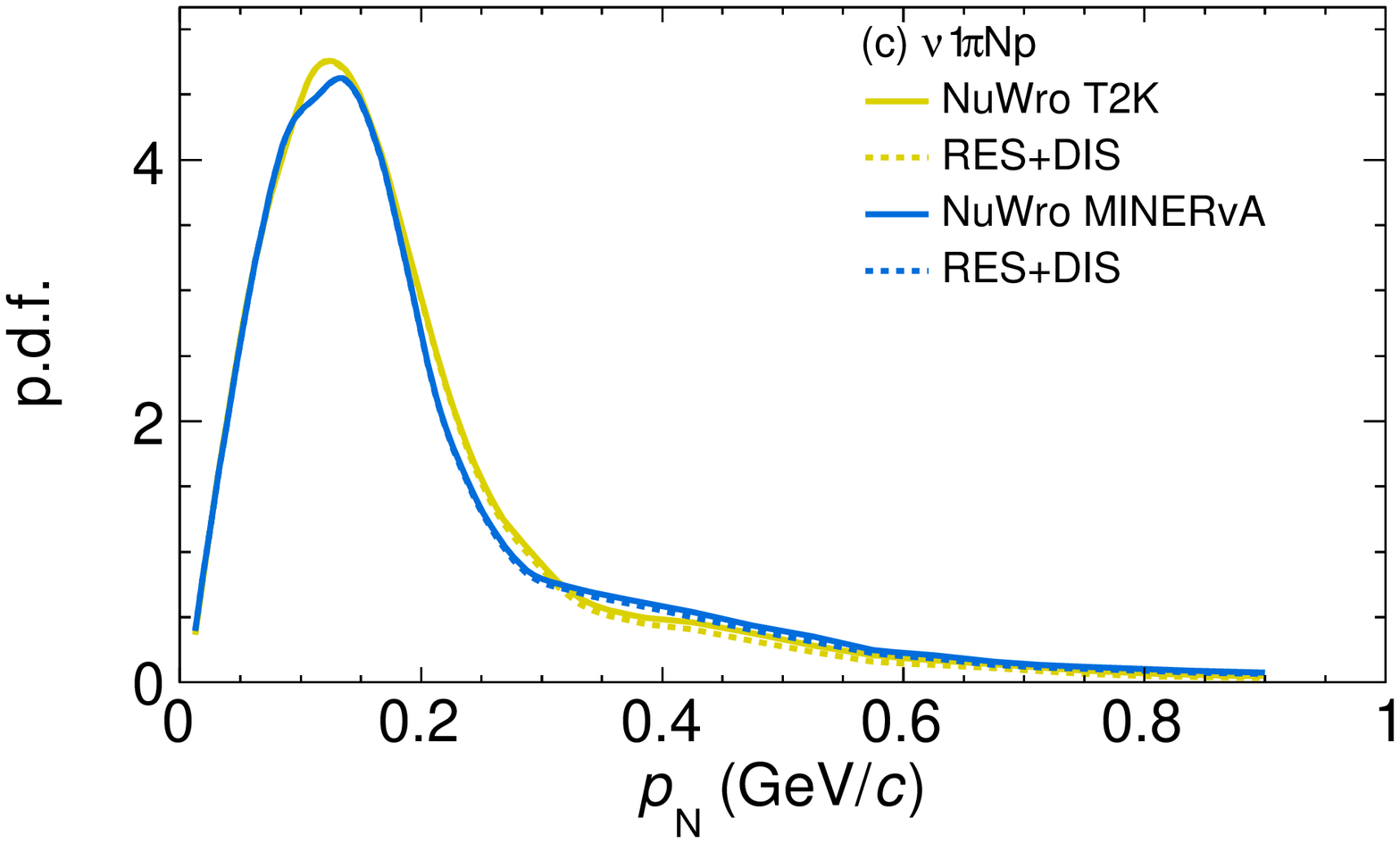}\\
\includegraphics[width=\figwid\columnwidth]{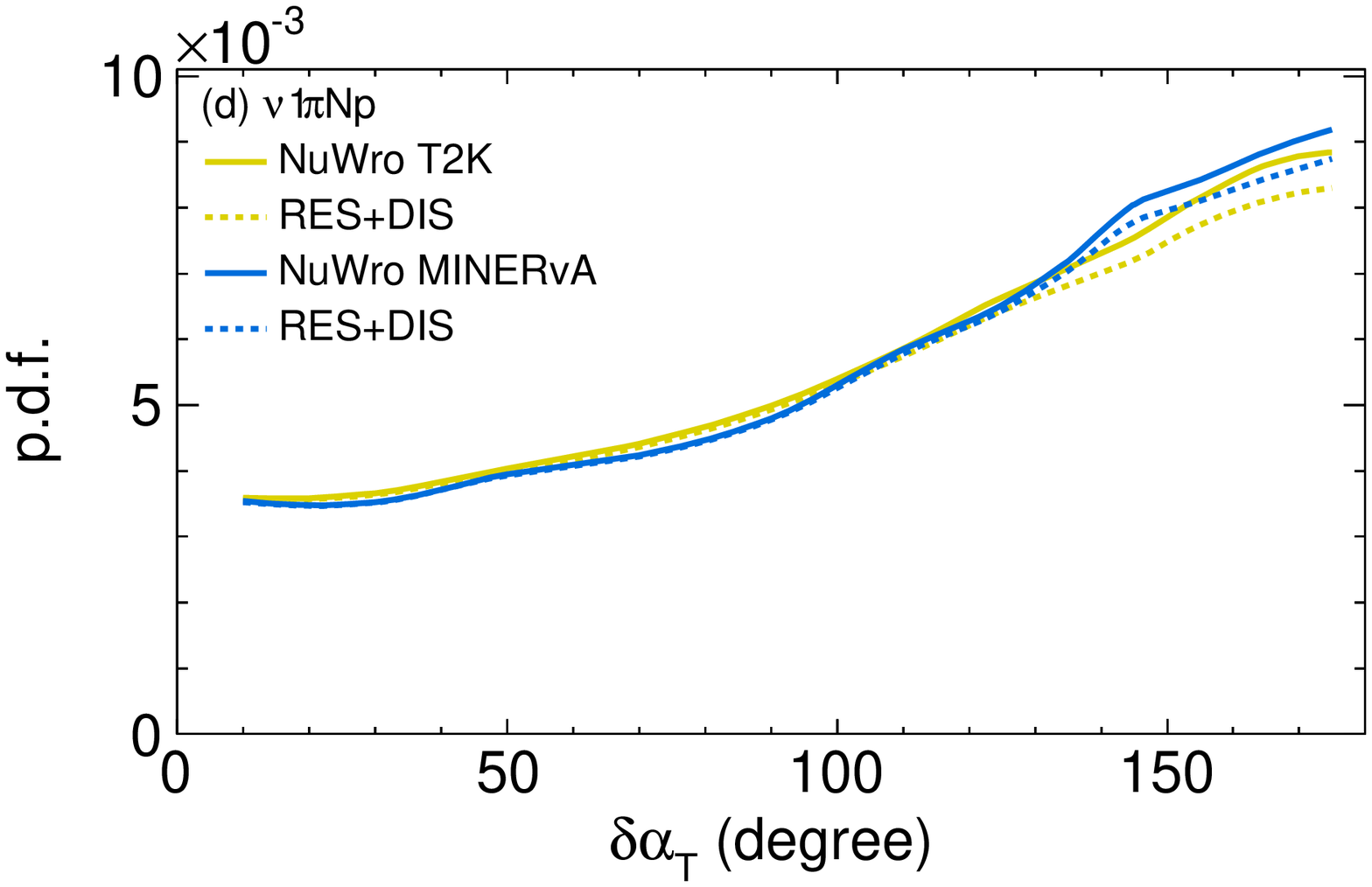}
\caption{Model comparisons for $\nuonepi$ in the same layout as in Fig.~\ref{fig:nubarzeropi}. The RES+DIS components are shown. }\label{fig:nu1pi}
\end{figure}

\begin{figure}[!ht] 
\centering
\includegraphics[width=\figwid\columnwidth]{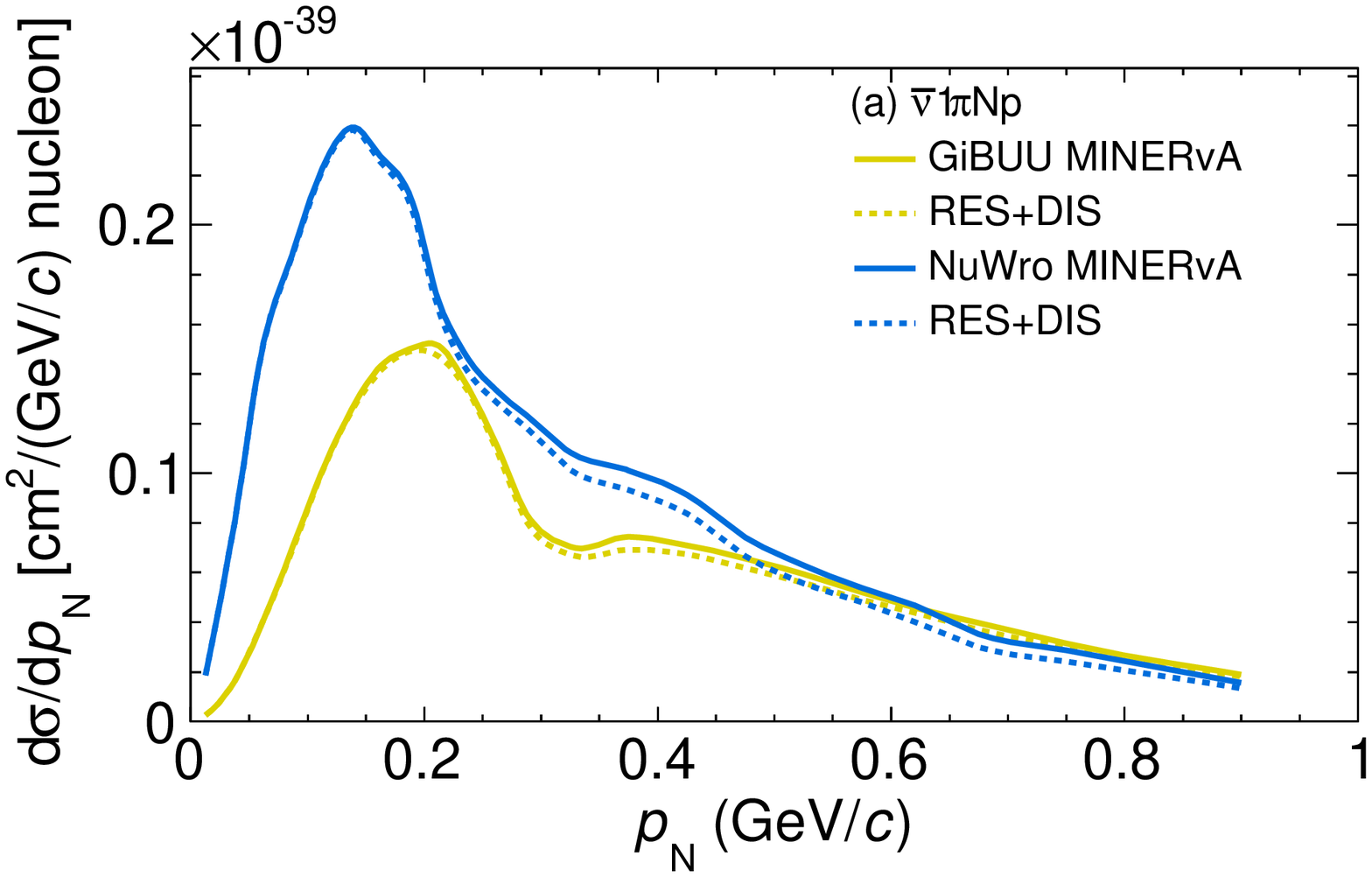}\\
\includegraphics[width=\figwid\columnwidth]{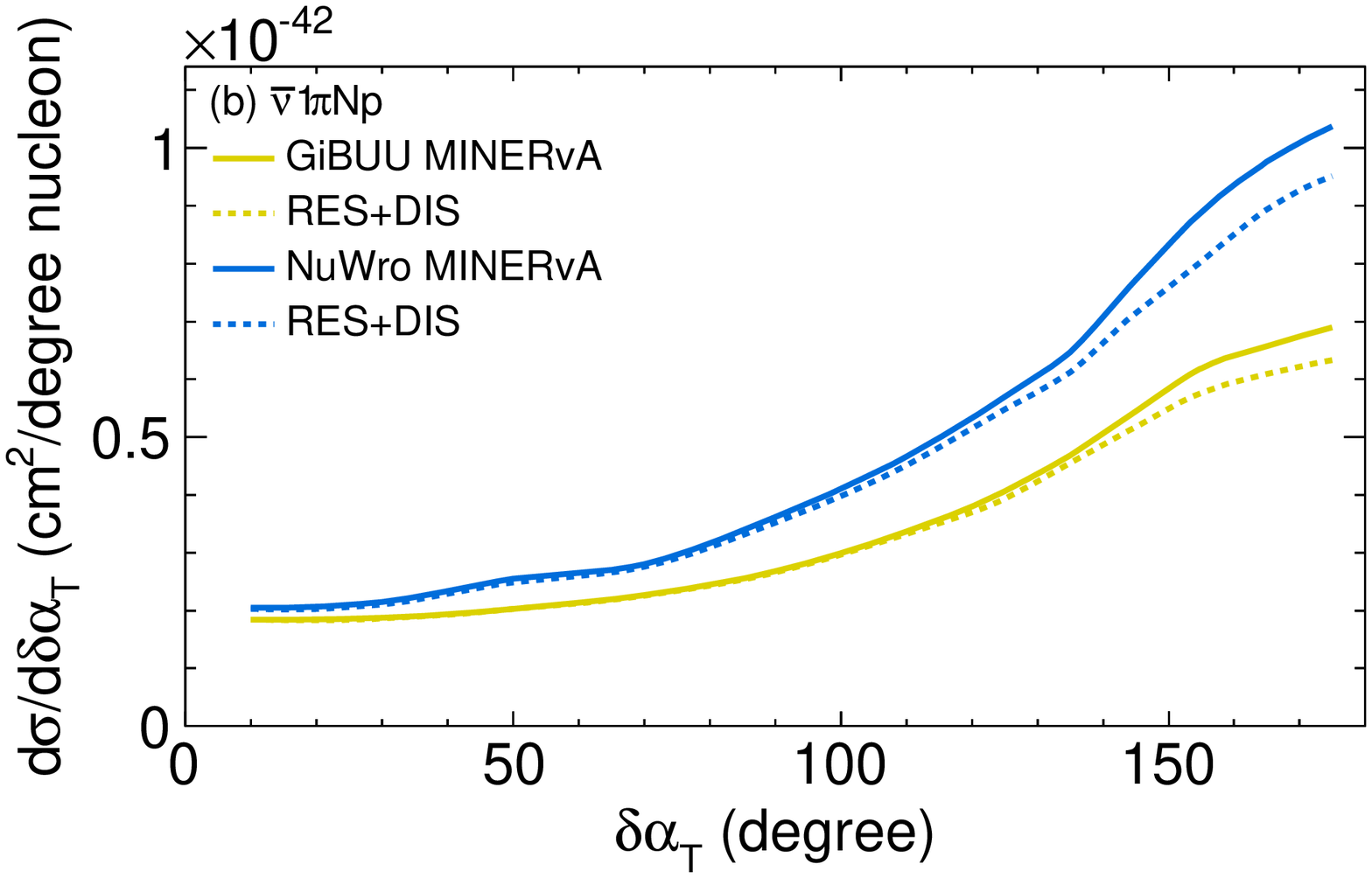}\\
\includegraphics[width=\figwid\columnwidth]{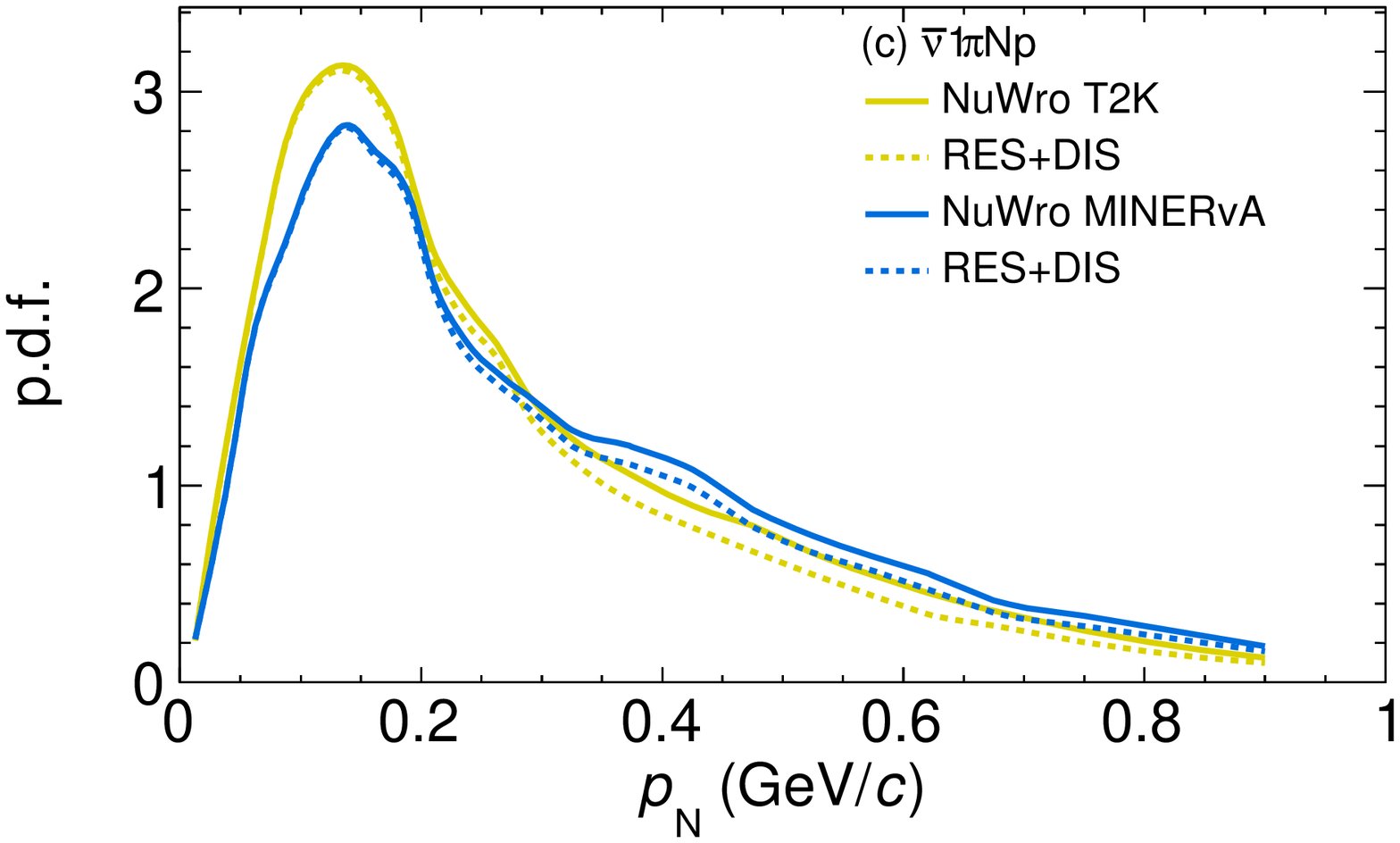}\\
\includegraphics[width=\figwid\columnwidth]{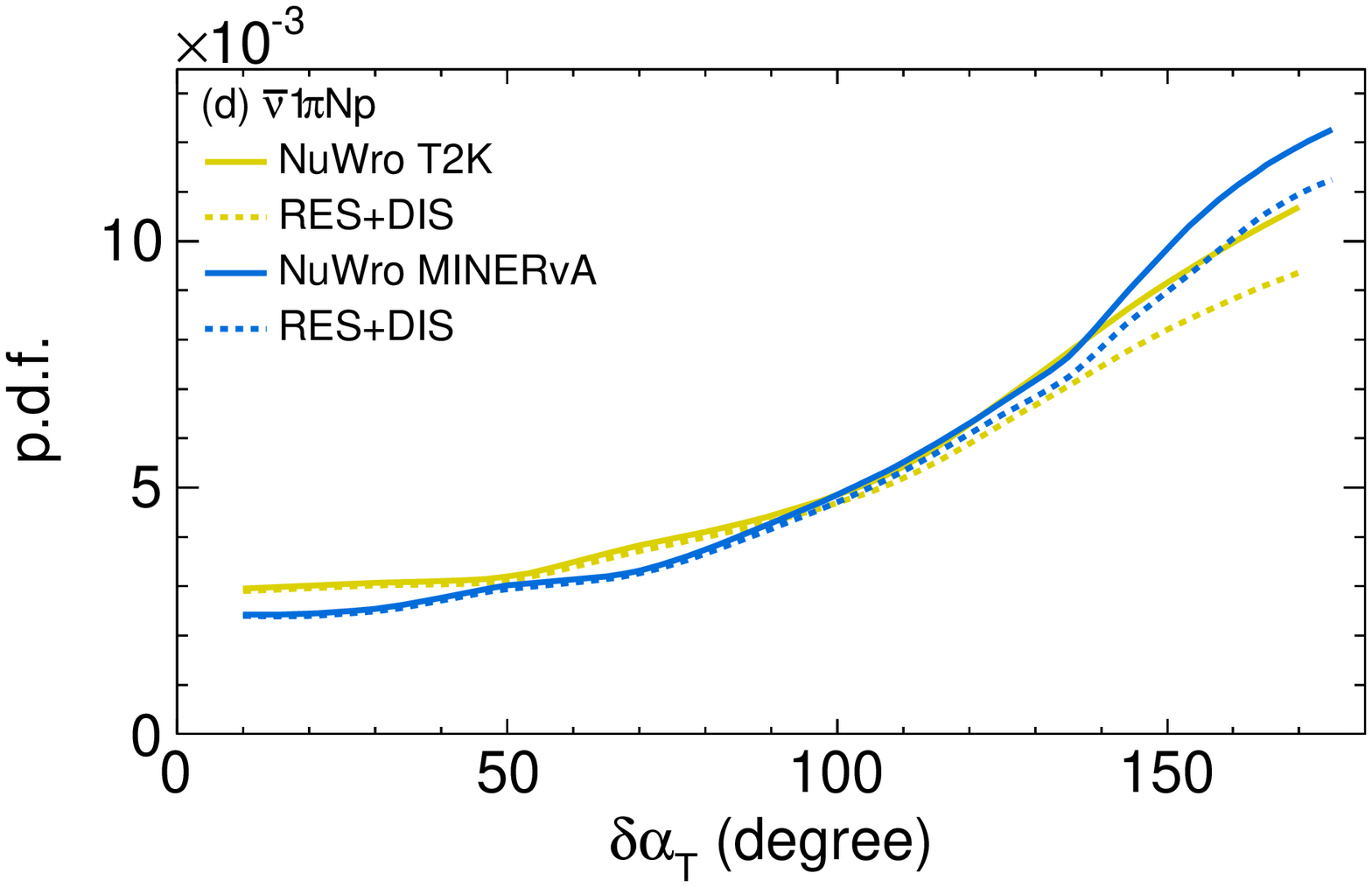}
\caption{Model comparisons for $\nubaronepi$ in the same layout as in Fig.~\ref{fig:nu1pi}.} \label{fig:nubar1pi}
\end{figure}

The channels $\nuonepi$ and $\nubaronepi$ (Figs.~\ref{fig:nu1pi} and~\ref{fig:nubar1pi})  are dominated by RES+DIS contributions. 
The variable $\pn$ is defined in such a way that, for events with $\Delta(1232)$ excitation  decaying into a charged pion and a proton but not suffering from FSI effects, it is equal to the initial-nucleon momentum. This time the target nucleon is a proton and the difference between $\nuonepi$ and $\nubaronepi$  is in the charge of the final-state pion.
In both cases, a clear Fermi motion peak is predicted by  GiBUU and NuWro, but again with different shapes. This peak  provides the most direct constraint on the Fermi motion of the initial-state proton, which has not been yet directly studied in neutrino interactions.  The overall amount of RES+DIS events without FSI is   similar in both models, as demonstrated by the cross section at $\dat\rightarrow0$.  And yet, the $\dat$ rising trend indicates the different FSI strength in the two models. Also, by comparing the rising trend of $\dat$ between the two channels, one can conclude that  in both models the $\nubaronepi$ channel suffers  stronger FSI, making a higher   tail in $\pn$.

The calculations show that, in the kinematic regions of $\nuonepi$ and $\nubaronepi$ probed by    MINERvA and T2K  experiments, the shape of $\pn$ and $\dat$ would depend only weakly  on the neutrino energy. This is a very strong statement to be verified by experiments. 

\section{Discussion and outlook}\label{sec:discussion}

The next-generation  long-baseline neutrino oscillation experiments DUNE and Hyper-Kamiokande  aim to measure CP violation based on a comparison of the neutrino and antineutrino  oscillation patterns. To achieve this goal a very good control of nuclear effects in (anti)neutrino scattering is necessary. In particular, it is very important to control nuclear effects that are inherently distinct for neutrino and antineutrino    scattering. The proposed generalized final-state correlations among the charged lepton and hadrons are minimally affected by nucleon-level phenomena and the beam flux~\cite{Lu:2015tcr}. They directly reveal details of the nuclear effects and allow to test theoretical models.

In water Cherenkov detectors one looks mainly at CCQE events and it is critical to analyze the oscillation signal with a model describing precisely the distributions of the charged-lepton kinetic energy and angle. Recent studies~\cite{Alvarez-Ruso:2017oui, Betancourt:2018bpu, Bodek:2018lmc}  have shown the effects of  the initial states  in measuring the oscillation parameters. In the $\delta_\textrm{CP}$  measurements where charged-lepton kinematics are used to infer the neutrino energy, understanding of the underlying neutron and proton Fermi motion is particularly important. 
Because the observed Fermi motion is  weighted by the underlying (anti)neutrino-nucleon cross section, a direct measurement of proton Fermi motion in the  $\nuonepi$ and $\nubaronepi$ channels using $\pn$ could provide valuable knowledge of the response of the constituent proton to the different electroweak probes mediated by $W^\pm$ bosons, which could mimic CP violation in neutrino oscillations.

It would be interesting to use the proposed observables in LAr experiments. Argon is a heavier nuclear target which makes nucleon FSI effects stronger. However, in LAr detectors lower-momentum knocked-out protons with weaker FSI effects are also reconstructed. In the recent measurement of electron scattering on an argon target~\cite{Dai:2018gch}, the Fermi motion peaks from carbon and argon are shown to have different shapes. While Fermi motion of the constituent nucleons in argon nuclei can be inferred with electron scattering, it can be determined \textit{in situ} in neutrino interactions by measuring $\pn$ in the $\nuzeropi$, $\nuonepi$, and $\nubaronepi$ channels, the response to the axial current elicited.

As is suggested by the MINERvA $\nuzeropi$  measurement~\cite{Lu:2018stk}, state-of-the-art generators fail  in the transition region of  $\pn$  between the Fermi motion peak and the region that is dominated by FSI and 2p2h. Without the Fermi motion peak in the $\nubarzeropi$ channel, the source of this model deficit could be determined. More importantly, as Ref.~\cite{Lu:2018stk} suggests that this intermediate $\pn$ region is where the MINERvA empirical 2p2h enhancement~\cite{Rodrigues:2015hik, Gran:2018fxa} is strongest, one might suspect that the deficit could be related  to the modeling of 2p2h. In the $\nubarzeropi$ channel, the ambiguity caused by the Fermi motion tail in constraining 2p2h models is removed,  potentially allowing a  better understanding of this complicated mechanism. As such, the interplay of several dynamical mechanisms---the initial state, QE, RES, and 2p2h interactions, and nucleon and pion FSIs---could be resolved. 

It is important to note that, for determination of the initial state,   neutral-pion production can also be considered:
\begin{align}
~\nu\nucleus&\to\lepton^-\proton\pi^0\fshadron,\label{eq:chanE}
\end{align}
with a  nucleon-level interaction,
\begin{align}
\nu\ \neutron&\to\lepton^-\ \proton\pi^0.
\label{eq:chanNE}
\end{align}
The generalized final-state correlations from Eq.~(\ref{eq:chanE}) have similar sensitivity to  Fermi motion as from the channel in Eq.~(\ref{eq:chanA}). In this paper we focused on the channels with charged pions in the final state, but a measurement in this channel could provide complementary information.  

Apart from nuclear-effect measurements, by selecting the Fermi motion peak in the $\pn$ distribution one can select a high-purity sample of genuine QE and RES events that do not experience FSIs in the $0\pi$ [Eq.~(\ref{eq:chanA})] and $1\pi$ [Eqs.~(\ref{eq:chanC}-\ref{eq:chanD})] channels, respectively. In such samples the neutrino energy  can be precisely reconstructed, as was first illustrated for the QE events in Ref.~\cite{Furmanski:2016wqo}.

The generalized final-state correlations focus on kinematics imbalances in exclusive reactions, which can be complemented by calorimetric inclusive measurements around the vertex region~\cite{Rodrigues:2015hik, Gran:2018fxa}. For example,  a better $\pn$ peak measurement could be achieved by  imposing a cut on the vertex energy, so that events  other than no-FSI QE/RES  are removed from the $0\pi$/$1\pi$ channel(s). 

Finally, the selection of three charged particles required in the  $\nuonepi$ and $\nubaronepi$   channels has important experimental implications. As was  first proposed and discussed in Refs.~\cite{Lu:2015hea, Lu:2015vri},
it enables to extract, on an  event-by-event basis, neutrino- and antineutrino-hydrogen interactions from compound  targets that contain hydrogen atoms. In addition to the  double-TKI~\cite{Lu:2015hea}, the single-TKI (imbalance between the charged lepton and  hadrons) and $\pn$ in principle could also provide separation power between interactions on hydrogen and heavier nuclei when the detector responses are optimized.

\begin{acknowledgments}
We thank Trung~Le for  the discussions about  pion secondary interactions in   MINERvA, and  Ulrich~Mosel for helpful comments on the manuscript.  J.T.S. was supported by NCN Opus Grant No.~2016/21/B/ST2/01092, and also by the Polish Ministry of Science and Higher Education, Grant No.~DIR/WK/2017/05. X.L. was supported by the UK Science and Technology Facilities Council. X.L. thanks Kevin~McFarland and  Ulrich~Mosel  for valuable discussions.

\end{acknowledgments}

\bibliographystyle{apsrev4-1}

\bibliography{bibliography}

\end{document}